# Embedded Systems Architecture for SLAM Applications

Jie Tang, Shaoshan Liu, and Jean-Luc Gaudiot, Fellow, IEEE

**ABSTRACT --** In recent years, we have observed a clear trend in the rapid rise of autonomous vehicles, robotics, virtual reality, and augmented reality. The core technology enabling these applications, Simultaneous Localization And Mapping (SLAM), imposes two main challenges: first, these workloads are computationally intensive and they often have real-time requirements; second, these workloads run on battery-powered mobile devices with limited energy budget. In short, the essence of these challenges is that performance should be improved while simultaneously reducing energy consumption, two rather contradicting goals by conventional wisdom. In this paper, we take a close look at state-of-the-art Simultaneous Localization And Mapping (SLAM) workloads, especially how these workloads behave on mobile devices. Based on the results, we propose a mobile architecture to improve SLAM performance on mobile devices.

## 1. INTRODUCTION

In recent times, we have observed a clear technology trend: the rapid rise of autonomous vehicle, robotics, virtual reality (VR), and augmented reality (AR) applications, and we expect the explosive growth of these applications to continue in the future. Robotics applications require the robot to sense the environment, and be able to identify its own location at real time. In VR, users would like to interact with objects in the virtual environment without using external controllers. In AR, the object being rendered needs to fit in the real-life 3D environment, especially when the user moves. Even more importantly, in autonomous vehicles, such as drones, the vehicle must find out its location in a 3D environment. The core technology enabling these applications is Simultaneous Localization And Mapping (SLAM), which constructs the map of an unknown environment while simultaneously keeping track of the location of the agent. It is a complex pipeline that consists of many computation-intensive stages, each performing a unique task.

The rise of these new applications impose two main challenges: first, these workloads have complex pipelines and are computation intensive and they often have tight real-time requirements. For example, in SLAM, sensor data can rush in at a rate as high as 1 KHz, meaning that the computation pipeline needs to process sensor data to produce 1,000 position data in a second, it also means that the longest stage of the pipeline cannot take more than 1 ms to process. In addition, the samples form a time series such that the samples cannot be processed in parallel. Second, these workloads could run on battery-powered mobile devices with extremely limited energy budget. Many VR and AR applications run on mobile phones today and drain battery quite rapidly. For example, our experiments showed that the Google Tango device, when running AR applications, would see its battery drain within forty minutes.

In this paper, we implement a SLAM system on existing mobile devices, and we examine its performance and energy consumption. Based on our initial findings, we propose a SLAM architecture to further improve the performance and energy efficiency of these applications on mobile devices.

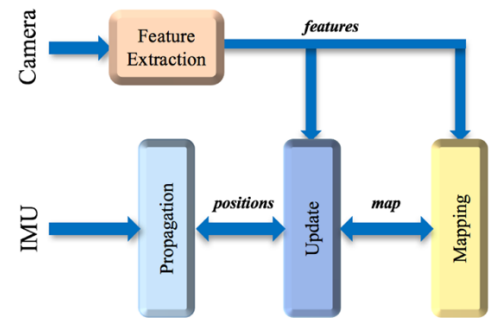

**Fig 1. simplified SLAM pipeline**

## 2. SLAM

SLAM is the problem of constructing or updating a map of an unknown environment while simultaneously keeping track of an agent's location within it [1]. Fig. 1 shows a simplified version of the general SLAM pipeline which operates as follows:

1. The *Inertial Measurement Unit*, or IMU, consists of a gyroscope to measure angular velocity and accelerometers to measure acceleration in the three axis. The IMU produces six data points (angular velocities in three different axis and accelerations in the three axis) at a high rate and feeds the data to the propagation stage.

2. The main task of the *Propagation Unit* is to integrate the IMU data points and produce a new position. Since IMU data is received at fixed intervals, by integrating the accelerations twice over time, we can derive the displacement of the agent during the last interval. However, since the IMU hardware usually has bias and inaccuracies, we cannot fully rely on Propagation data, lest the positions produced gradually drift from the actual path [2].

3. To correct the drift problem, we use a camera to capture frames along the path at a fixed rate, usually at 60 FPS.

4. The frames captured by the camera can be fed to the *Feature Extraction Unit*, which extracts useful corner features and generates a descriptor for each feature [3].

5. The features extracted can then be fed to the *Mapping Unit* to extend the map as the Agent explores [4]. Note that by map, we mean a collection of 3D points in space, each 3D point would correspond to one or more feature points detected in the *Feature Extraction Unit*.

6. Also, the features detected would be sent to the *Update Unit* which compares the features to the map [3]. If the detected features already exist in the map, the Update unit can then derive the agent's current position from the known map points. By using this new position, the Update Unit can correct the drift introduced by the Propagation Unit. Also, the Update unit updates the map with the newly detected feature points.

## 3. SLAM IMPLEMENTATION ON MOBILE ARCHITECTURE

In this section we implement SLAM on mobile devices and evaluate its performance and power consumption.

### 3.1 Existing Mobile Architecture

In our experiments, we implemented our SLAM system on a ARM v8 mobile SoC [5]. Figure 2 presents the general design of mobile SoC: the SoC consists of a four-core CPU, running at 2 GHz, the four cores share a L2 cache. Besides the CPU, the SoC consists of a DSP, and a GPU, and the CPU communicates with the DSP and the GPU through shared memory. Note that the I/O subsystem is directly connected to the CPU. If acceleration is needed, the CPU would then dispatch workloads to DSP and GPU through shared memory.

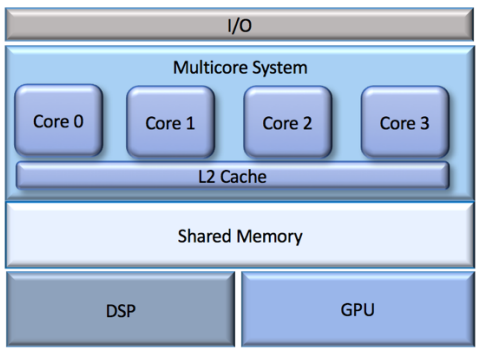

Fig 2. existing mobile architecture

Before implementing the SLAM system onto this mobile SoC, we first tried to understand the power consumption behavior of each computing unit. We implemented several stress test to stress CPU, DSP, and GPU respectively and the results are shown in Figure 3: at its peak, each CPU consumes about 2.5 W, DSP consumes about 1.5 W, and GPU consumes about 2.3 W. This implies that if a DSP delivers similar performance compared to a CPU core, it would be more energy-efficient to use DSP.

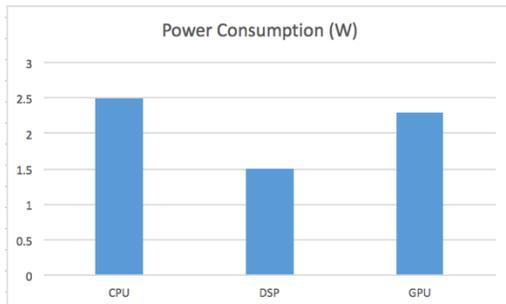

Fig 3. power consumption analysis

### 3.2 SLAM Implementation on Mobile Architecture: CPU

As a first step of implementing the SLAM system on mobile SoC, we implemented a CPU-only version. As shown in Figure 4, we implemented the feature extraction thread on core 0, the propagation thread on core 1, the update thread on core 2, and the mapping thread on core 3. These different threads communicate with each other through the shared memory.

A typical flow works as follows: first, image comes in and triggers the feature extraction thread, which then extracts feature points from the target image. Once features get extracted, the update thread and the mapping thread would pick up these features from shared memory. In addition, the propagation thread consumes IMU samples and sends propagation results to the update thread through shared memory.

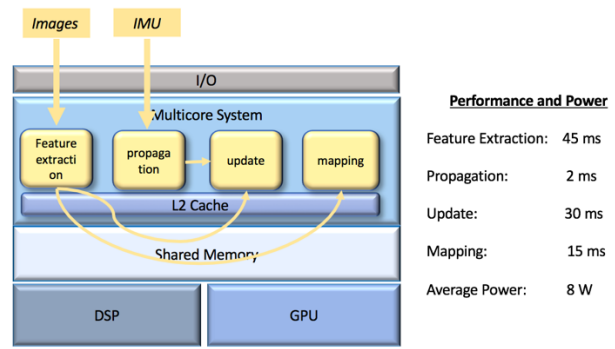

Fig 4. CPU implementation on mobile architecture

Delving into its performance, we found that the feature extraction stage takes 45 ms to finish, and the propagation stage takes 2 ms, the update stage takes 30 ms, and the mapping stage takes 15 ms. On average, the overall system consumes 8 W when running this system. Using this system, we could only consume 15 to 20 images per second, and about 200 IMU samples per second. Ideally, for the SLAM system to meet real-time requirements, we would like to have it consume 30 images per second.

### 3.3 SLAM Implementation on Mobile Architecture: Heterogeneous Computing

Next we look at what we could do in the software layer to improve performance. We started by examining which stage could be easily parallelized: propagation and update stages both rely on a time-series of data, such that the new data has dependency on the previous data, and thus it is not straightforward to parallelize these two threads. So as the mapping thread, which extends the existing map based on new incoming data. These tasks are not suitable for parallel execution and thus not fit for GPU or DSP. That leaves us with the bottleneck of the pipeline, which is the feature extraction thread.

As shown in Figure 5, we implemented our feature extraction stage on CPU, GPU, and DSP. In terms of performance, feature extraction takes 45 ms, 50 ms, and 20 ms respectively on CPU, GPU, and DSP. Note that using GPU even degrades performance, due to the high overhead to setup data and computing resources before we could start computation. In terms of energy consumption, each execution of feature extraction consumes 112 mJ, 115 mJ, and 30 mJ respectively on CPU, GPU, and DSP. Therefore, DSP seems a perfect fit for feature extraction, reducing execution time by half and improving energy efficiency by almost four-fold.

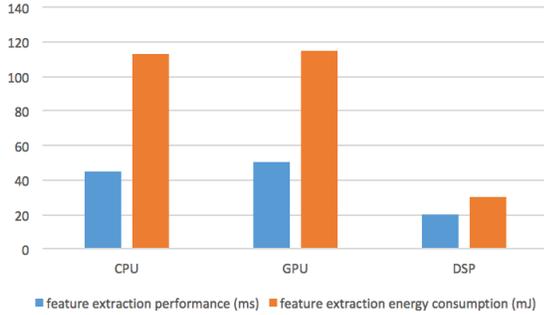

**Fig 5. feature extraction performance and power**

With DSP acceleration, let us examine the overall system performance. As shown in Figure 6, although we utilize DSP for feature extraction, we still require a core to relay data from I/O to DSP since DSP is not directly connected to the I/O system. This introduces two problems: first, we still need to use a CPU thread to relay data, incurring extra power consumption. This is reflected in the average system power consumption, which increased to 9 W from 8 W due to the involvement of DSP. Second, this design has its impact on performance as well. As each time an image is received by the CPU, it needs to allocate an extract copy of it in memory then sends it to the DSP. By our measurement, this would take between 1 ms to 3 ms. Even worse, in a managed runtime environment, this could very likely introduce garbage collections [6], which freezes the whole system for over 100 ms, and thus stalling the whole SLAM pipeline, leading to the system losing track of itself.

Nonetheless, the SLAM system could now process 30 FPS, resulting in real-time performance, though this real-time performance could be interrupted by buffer copying and garbage collection, and the power consumption is not ideal.

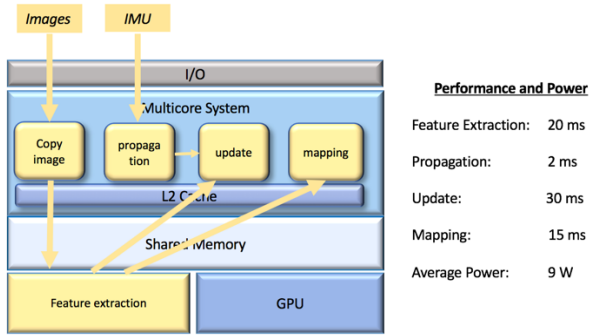

**Fig 6. heterogeneous implementation on mobile architecture**

## 4. ARCHITECTURE FOR SLAM
Based on the initial findings presented above, we propose an Architecture to optimize the performance and energy consumption for SLAM workloads.

### 4.1 Direct IO for DSP
As discussed in the previous section, in current implementations, the DSP is not directly connected to the IO system, and a CPU core has to act as a relay to copy image data in memory for the DSP to consume. Each incoming image is at least 3 MB, at 30 FPS, this means 100 MB of memory allocation per second. This not only wastes CPU resources, leading to extra power consumption, but also leads to memory problems, such as memory copy overheads and garbage collections.

To solve this problem, we propose direct IO for DSP such that the DSP gets directly connected to the image sensor IO pins, allowing the DSP to directly consume image data without the involvement of a CPU core. This way we free up one CPU core, and gets rid of the extra memory copying.

Some may worry that this design would introduce extra power consumptions. To find out the power overhead of driving the extra I/O pins, we implement a test to repeatedly read data from a serial IO pin and measured its power consumption overhead, and the measured power consumption driving the pins is only 0.1 W.

### 4.2 ScratchPad Feature Buffer
With DSP Direct IO, we could offload feature extraction tasks to the DSP, however, the extracted features would go into memory, and later on the threads running on CPU cores would make a copy of the features then consume them. On mobile devices, the memory access latency is at the scale of 100 ns. To consume the extracted features, each second the CPU threads make at least 6000 memory access requests. Based on our profiling results, the extracted feature memory accesses account for about 20% of the total execution time of the update thread and the mapping thread.

From our initial study, we understand that from each image we generate at most 4 KB of feature data. So if could design a low-latency ScratchPad memory [7] to directly feed the extracted features to the CPU cores, then we could significantly improve overall SLAM performance.

To understand the performance and power consumption of such ScratchPad memory, we utilize CACTI to analyze our design [8]. In this design, the size of the ScratchPad memory is 8KB with two memory banks, one for each batch of image features, this way we leave enough swap space to store the features from two frames, thus making sure we do not have dropped frames if the CPU cores temporarily lag behind the DSP, and the ScratchPad memory should be implemented in at least 32 nm technology. The results are summarized in Table 1, this ScratchPad buffer would consume at most 0.15 W when active and it has a leakage power of only 0.002 W. In addition, each memory access now takes only 0.4 ns instead of 100 ns. Using this buffer, we would significantly improve the performance of the SLAM pipeline, by at most 20%.

**Table 1. ScratchPad feature buffer performance and power**

| Access Time (ns) | Dynamic Power (W) | Leakage Power (W) |
|---|---|---|
| 0.4 | 0.15 | 0.002 |

### 4.3 SLAM Trigger
Once the extracted features get written to the ScratchPad memory, we need a mechanism to notify the consumers, including the update thread and the mapping thread. Instead of having these two threads busy-waiting for the update, we implement a notification mechanism, SLAM Trigger, and it works as follows: in the Feature Buffer controller we have a register to store the current filled feature memory bank ID. Once a bank is filled, the bank number is written in this register and the Feature Buffer controller would then interrupt the CPU to notify the update thread and the mapping thread about the incoming new features. Once these two threads are woken up, the CPU would lock the filled bank and then clear the filled feature memory bank ID while the DSP is writing to the other bank. Once the CPU finishes consuming the feature buffer bank, it releases the bank and the threads go back to sleep. By swapping the two banks between the consumer and the producer, we effectively synchronize the two parties. If the consumer lags behind the producer, we can throttle the image frame rate to allow the consumer to catch up.

### 4.4 Putting Everything Together
Now let us incorporate these designs into a new architecture optimized for SLAM workloads. As shown in Figure 7, DSP is connected to the image sensor through Direct IO such that each time a new image comes in, it triggers the feature extraction task on DSP, and the extracted feature gets written to the ScratchPad memory. Once the extracted features fill a ScratchPad bank, the ScratchPad controller interrupts the CPU to trigger SLAM.

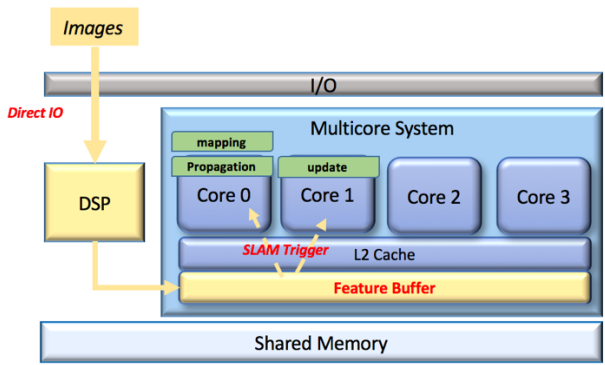

**Fig 7. SLAM mobile architecture**

In this design, we have both the propagation thread and the mapping thread running on Core 0, and the update thread sharing Core 1. This is made possible since now we use a trigger mechanism for the mapping thread, such that the mapping thread is only active when data is available. When the mapping thread is active, the incoming IMU data get buffered and later the propagation thread would consume the IMU samples in batch. We have verified that this technique does not impact the performance and accuracy of the SLAM system.

### 4.5 Performance and Power Consumption
To estimate the performance and power consumption of such system, we use both simulation and real measurements. First, to understand the power requirements, we implement the update thread, the mapping thread, and the propagation thread on two CPU cores, and we implement the feature extraction task on DSP and measure the system power consumption. On average, this system consumes about 5 W. Including the extra power required for the feature buffer and driving the extra IO pins, the system power consumption should be at about 5.3 W.

**Table 2. performance estimation with the SLAM architecture**

| Feature Extraction (ms) | Propagation (ms) | Update (ms) | Mapping (ms) |
|---|---|---|---|
| 20 | 2 | 24 | 12 |

For performance, as shown in Table 2, we measure that it takes 20 ms to run feature extraction on DSP, and we have profiled that feature memory access takes about 20% of the execution time of the update and mapping threads. By dropping memory access time from 100 ns to about 0.4 ns, we basically eliminate the feature access overheads, thus improving the update thread and mapping thread performance by 20%. With this SLAM architecture, the device consumes about 5 W and is able to process almost 50 FPS of image, generating accurate real-time position updates.

## 5. CONCLUSIONS
SLAM is the core technology enabling the new generation of applications, including autonomous robots, AR, VR. In this paper, we have implemented a SLAM system on existing mobile devices, and we have examined its performance and energy consumption. Based on our initial findings, we propose a SLAM architecture which is able to process 50 FPS of incoming images at 5 W, compared to 15 to 20 FPS at 8 W in state-of-the-art mobile systems. Note that this is only an initial attempt in designing a SLAM mobile device. In the next step, we examine how to further improve the performance of SLAM, especially the update and feature extraction tasks, and how to further reduce power consumption.

## 6. ACKNOWLEDGMENTS
This work is partly supported by the National Science Foundation under Grant No. and XPS-1439165. Any opinions, findings, and conclusions or recommendations expressed in this material are those of the authors and do not necessarily reflect the views of NSF.